# Structured illumination microscopy for dual-modality 3D sub-diffraction resolution fluorescence and refractive-index reconstruction

Shwetadwip Chowdhury[1], Will J. Eldridge[1], Adam Wax[1], and Joseph A. Izatt[1,*]

**Authors information:**
[1]Duke University, Biomedical Engineering Department
1427 FCIEMAS, 101 Science Drive, Box 90281
Durham NC, 27708

**\*Corresponding Author:** joseph.izatt@duke.edu

## Abstract

Though structured illumination (SI) microscopy is a popular imaging technique conventionally associated with fluorescent super-resolution, recent works have suggested its applicability towards sub-diffraction coherent imaging with quantitative endogenous biological contrast. Here, we demonstrate that SI can efficiently integrate the principles of fluorescent super-resolution and coherent synthetic aperture to achieve 3D dual-modality sub-diffraction resolution, fluorescence and refractive-index (RI) visualizations of biological samples. Such a demonstration can enable future biological studies to synergistically explore molecular and biophysical/biochemical mechanisms at sub-diffraction resolutions. We experimentally demonstrate this framework by introducing a SI microscope capable of 3D sub-diffraction fluorescence and RI imaging, and verify its biological visualization capabilities by experimentally reconstructing 3D RI/fluorescence visualizations of fluorescent calibration microspheres as well as alveolar basal epithelial adenocarcinoma (A549) and human colorectal adenocarcinmoa (HT-29) cells, fluorescently stained for F-actin.

# I. Introduction

Optical microscopy is an indispensable tool in the biological sciences, and allows one to visualize sub-micron biological features with various forms of imaging contrast. However, though mechanisms of image contrast can be chosen based on specific experimental aims and considerations, the imaging resolution is still generally constrained by the optical diffraction limit, which can preclude visualization of key biological components at sub-diffraction size scales. Investigations into visualizing such features has led to the development of a host of sub-diffraction resolution microscopy techniques that utilize a variety of creative mechanisms to image features beyond the diffraction limit.

These techniques are typically separated into two classes based on the type of samples they are used to image. The first class focuses on achieving sub-diffraction resolution imaging in coherently diffracting samples, where the imaging process can be modelled via information theory to have an invariant number of degrees-of-freedom (one of which is imaging resolution). By sacrificing degrees of lesser importance, one can in principle improve the imaging resolution to beyond the classical diffraction limit [1-3]. Of closest relation to this work, synthetic aperture (SA) is a subset of coherent sub-diffraction resolution microscopy that exploits temporal degrees-of-freedom by using multiple angled illuminations to spatiotemporally encode more frequency content into images than conventionally allowed by the system's physical aperture. SA has become a popular coherent sub-diffraction microscopy tool in the biological sciences and has enabled high-speed, endogenous-contrast biological imaging at both high resolutions and high throughputs. Furthermore, it has shown applications for speckle reduction, non-interferometric phase-retrieval, and tomographic imaging [4-9].

The second class of techniques, often (but not exclusively) referred to as "super-resolution", is targeted towards fluorescence-emitting samples, where the tagged fluorophores are usually designed to either deplete or photoswitch. For fluorophores with depletion properties, stimulated emission depletion (STED) is a popular super-resolution technique that synchronizes two scanning excitation/depletion beams to directly minimize the extent of a scanning excitation point via saturated stimulated emission [10, 11]. For fluorophores with photoswitching properties, photo-activated localization microscopy (PALM) is a popular super-resolution choice and operates by exciting random sparse subsets of fluorophores in a single acquisition to resolve individual fluorophores at sub-diffraction resolutions [12]. It then repeats this process across multiple acquisitions to aggregate together the total number of localizations into a final super-resolved image. Both STED and PALM are now established tools in fluorescence microscopy and routinely allow molecular-specific biological imaging at resolutions well below 100 nm.

Fluorescence microscopy is the standard choice when imaging cells with molecular-specific and background-free imaging contrast, and has enabled visualization of organelle structures, protein interactions, gene expression, and cellular dynamics [13-17]. Conversely, coherent-diffraction microscopy is a popular choice to image samples with endogenous contrast and to quantitatively measure intrinsic biological parameters, such as dry mass, shear stiffness, and refractive-index/optical-path-length [8, 18-21]. Such parameters often directly relate to biophysical and biochemical properties of cells. As such, both fluorescence microscopy and coherent-diffraction microscopy have found considerable utility in biological imaging, and have facilitated complementary advances in biological analyses. Furthermore, addition of sub-diffraction resolution capabilities to both these classes of microscopies, individually, has significantly enhanced the ability to form biological conclusions, and has been shown by previous studies to directly enable specific important insights [10, 12, 22-28]. Unfortunately, because fluorescence and coherent-diffraction microscopies operate on fundamentally different image formation mechanisms, a sub-diffraction resolution technique applicable to both has remained elusive. This prevents users from synergistically combining the individual biological insights gained from these two regimes, which in turn can impede a comprehensive understanding of biological components and processes with important molecular and biophysical/biochemical properties.

Previous works, however, have demonstrated that structured illumination (SI) microscopy, already an established imaging technique for fluorescent super-resolution [29-34], has considerable utility as a SA technique when operated in the coherent imaging realm. Such studies have demonstrated SI's coherent sub-diffraction resolution imaging capabilities for visualizing 2D and 3D scattering/quantitative-phase (QP), as endogenous contrast mechanisms [35-43]. Hence, SI fills a niche role as a sub-diffraction resolution imaging solution suitable for studies probing molecular and biophysical/biochemical processes. To highlight this coherent/fluorescent multi-modal compatibility unique to SI, we recently introduced an imaging framework that utilized SI to enable 3D sub-diffraction resolution biological imaging of both QP and fluorescence [42]. Though clearly an important step towards our main goal of establishing a sub-diffraction imaging technique inclusive to multiple modalities, a valuable addition to the work would be to extend the SI framework to include 3D reconstructions of biological refractive-index (RI), the intrinsic optical parameter of biological samples giving rise to differences in QP. In pursuit of this, we recently demonstrated that coherent SI, as a SA technique, could be integrated into the framework for diffraction tomography (DT) and enable reconstructions of 3D refractive-index (RI) [44] (similar findings were also reported by [45], where a time-multiplexed implementation of SI was used for 3D RI reconstruction). In this current presented work, we extend these previous results to demonstrate a single optical system that uses SI to perform both 3D fluorescent and RI visualizations at sub-diffraction resolutions. This system is a potential solution towards multimodal, sub-diffraction visualization of biological components.

## II. 3D coherent and fluorescent transfer functions

It has been noted previously that image formation processes for fluorescent and coherent-transmission imaging, if combined with intensity and complex electric-field detection, respectively, can both be described with standard properties of linearity and space-invariance [46]. With this description, the fluorescent and coherent 3D diffraction-limited resolutions can directly be equated to the extent (i.e., support) of the 3D spatial-frequency transfer functions (TF). For coherent imaging, this TF takes the form of a spherical shell, with lateral ($k_x, k_y$) and axial ($k_z$) spatial-frequency magnitude bounds of $k_{c,\parallel} = \text{NA}/\lambda_c$ and $k_{c,\perp} = n(1 - \cos\theta)/\lambda_c$, respectively [47]. Here, λc is the wavelength used for coherent imaging, $n$ is the refractive index of the medium, $\theta$ is the maximum half-angle of light accepted by the detection aperture, and $\text{NA} = n\sin\theta$ is the system's numerical aperture. The TF for fluorescent imaging is known to be the autocorrelation of the coherent TF, and takes a 3D form similar to a torus, with lateral and axial spatial-frequency bounds of $k_{F,\|} = 2\,\text{NA}/\lambda_F$ and $k_{F,\perp} = n(1 - \cos\theta)/\lambda_F$, respectively [46]. Here, λF is the fluorescent emission wavelength. Coherent and fluorescent Abbe diffraction-limited lateral and axial resolutions (defined as the minimum-resolvable spatial-period) directly follow: $d_{c,\|} = 1/k_{c,\|} = \lambda_c/\text{NA}$, $d_{F,\|} = 1/k_{F,\|} = \lambda_F/2\text{NA}$, and $d_{F,\perp} = 1/k_{F,\perp} = \lambda_F/n(1-\cos\theta)$. Because the coherent TF has infinitesimal axial spatial-frequency extent, it does not have an associated Abbe axial resolution (i.e., the wavevector with axial component $k_{c,\perp}$ propagates through the whole coherent imaging volume and offers no optical sectioning) [48, 49]. Figs. 1(a,b) illustrate the coherent and fluorescent TFs. For sub-diffraction resolution imaging with both fluorescence and coherent transmission, the aim is to reconstruct an image with more spatial-frequency information than directly encompassed by the fluorescent and coherent TFs, respectively.

SI achieves this by illuminating spatial patterns onto the sample, which modulate the sample's sub-diffraction resolution information into the system's TF in the form of resolvable "beat" frequencies (i.e., Moiré patterns) [29]. Generally, this "beat" phenomenon occurs irrespective of imaging coherence, and thus can be utilized for sub-diffraction resolution imaging with both fluorescence and coherent-diffraction [42]. Interestingly, recent works have demonstrated that, when operated in the coherent imaging realm, SI's "beating" phenomenon theoretically decomposes to a coherent multiplex of sample diffraction from the individual angled illuminations composing the spatial illumination pattern [35, 42, 43]. Thus, via the principles of DT, SI can fill out 3D coherent k-space and reconstruct 3D RI by simply illuminating the sample with spatial patterns with incrementing spatial-frequencies [44]. We note that simply illuminating with the maximum magnitude spatial-frequency, as is typical in fluorescent SI, is not sufficient to fill out 3D coherent k-space due to the coherent TF's infinitesimal axial thickness. Fig. 1(c) illustrates the 2D and 3D filling of coherent k-space via SI with one angular orientation, as well as the 3D coherent k-space filling with 6 angular orientation (for isotropic lateral resolution). Fig. 1(d) illustrates the analogous 2D and 3D filling of fluorescent k-space for SI fluorescent imaging, with one and 3 angular rotations. We refer the reader to [31] for a more rigorous discussion about SI-enabled fluorescent super-resolution.

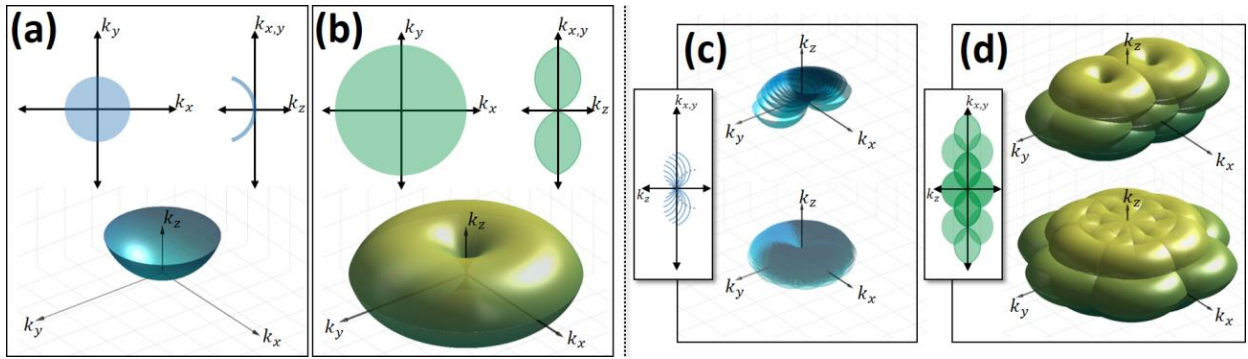

**Fig. 1.** Three-dimensional **(a)** coherent and **(b)** fluorescent diffraction-limited transfer functions are shown. **(c)** SI can fill out sub-diffraction regions in coherent 3D k-space by using incremented spatial frequencies, analogous to diffraction tomography. **(d)** SI can also fill out sub-diffraction regions in fluorescent 3D k-space [31]. Because the fluorescent transfer function shown in (b) is filled (not of infinitesimal axial thickness), SI spatial frequencies do not have to be incremented for fluorescent super-resolution, as is required in (c).

# III. Experimental methods and results

## A. Optical system design

We show in Fig. 2(a) a simplified schematic of an optical system designed for dual-modality 3D RI/fluorescence sub-diffraction resolution imaging. As described in previous works, this system used a single mode, broadband light source (NKT Photonics, EXW-6) [42, 44]. This light was spectrally filtered to 488 ± 12.5 nm (Semrock, FF01-482/25), and was used for both fluorescent excitation and coherent-diffraction imaging. The light was collimated from the illumination fiber source, passed through a 50:50 polarization beam splitter (PBS, Thorlabs, PBS251), and incident on an amplitude spatial light modulator (SLM, Holoeye, HED 6001). The SLM was programmed to display a sinusoidal pattern, which was imaged by a 4f system, consisting of a condenser lens (L1, AC508-200-A, Thorlabs) and illumination objective (OBJ, 63X, 1.4 NA Zeiss), onto the sample, which was mounted on an axial piezo-stage (MAX312D, Thorlabs). The excited fluorescence from the sample was collected in transmission mode by a second 4f system (with detection objective lens identical to the illumination objective), and directly imaged onto a camera (CMOS-F Pixelink). The coherent-transmission from the sample was collected by the same 4f system, but was spectrally separated from the sample fluorescence with a dichroic mirror (DM, Thorlabs, DMLP505). An image of the sample was coherently formed onto a Ronchi diffraction grating (RG, Edmund Optics, 72 lpmm), which incorporated copies of the image into 0 and ±1 diffraction-orders, which were transmitted through a third 4f system (L3 → L4), before being imaged onto a camera (CMOS-RI, Pixelink). An asymetric mask was inserted into the Fourier plane of the third 4f system to pass RG's +1st order, spatially filter the 0th order via pinhole (Edmund Optics, 52-869), and block all other diffraction orders. This enabled common-path off-axis interference with broadband illumination, which allowed recovery of electric-fields with reduced coherent noise and higher temporal stability. Figs. 1(b,c) show examples of raw fluorescent and coherent-interferogram acquisitions, respectively. Note the SI sinusoidal patterns that are clearly visible in both the fluorescent and coherent-diffraction acquisitions (indicated by yellow arrows in Figs. 1(b,d), respectively). Furthermore, Fig. 1(d) shows that the sinusoidal SI pattern is overlayed by the higher frequency off-axis interference pattern at the image-detection plane of CMOS-RI. This directly enables the sample's spatial frequency information to be separated from the conjugate and central ambiguity terms presented in the detected interferogram's Fourier transform, as is demonstrated in Fig. 1(e). Optical filtering of the sample's spatial frequency component (indicated in dashed yellow circle in Fig. 1(e)) yields the sample's complex electric-field, after inverse Fourier transform.

Because coherent SI essentially equates to a multiplexed version of SA, the retrieved electric-field from each individual SI pattern could be decomposed into the electric-field components arising from the individual angled plane-waves composing the SI pattern. These sample electric-fields were then processed by standard DT algorithms to reconstruct 3D sample RI. We emphasize here that, unlike conventional DT systems that require tilting mirrors, monochromatic laser illumination, and/or separate off-axis holography reference arms [8, 47, 50, 51], our system allows electric-field measurements from angled illuminations with a common-path, off-axis interferometer with no mechanically moving components and broadband laser illumination. This allows our RI reconstructions to benefit from increased temporal phase stabilities and decreased coherent noise artifacts [44].

Given a coherent illumination center-wavelength of $\lambda_c = 488$ nm, our expected coherent diffraction-limited lateral resolution was $d_{c,\parallel} = \lambda_c/\text{NA} \approx 350$ nm. For fluorescent imaging, our biological samples were tagged with Alexafluor 488 for F-actin visualization, which has the dominant peak-emission wavelength at 520 nm. Thus, our expected diffraction-limited lateral and axial fluorescent resolutions were $d_{F,\parallel} = \lambda_F/2\text{NA} \approx 185$ nm and $d_{F,\perp} = \lambda_F/n(1 - \cos\theta) \approx 550$ nm. After SI-enabled resolution enhancement, the expected lateral resolutions of the SI RI and fluorescent reconstructions were doubled to $d_{c,\parallel}/2 \approx 175$ nm and $d_{F,\parallel}/2 \approx 92$ nm, respectively. Expected axial resolutions of the SI RI and fluorescent reconstructions were $d_{c,\perp} \approx \lambda_c/n(1 - \cos\theta) \approx 508$ nm and $d_{F,\perp} = d_{F,\perp}/2 \approx 225$ nm, respectively.

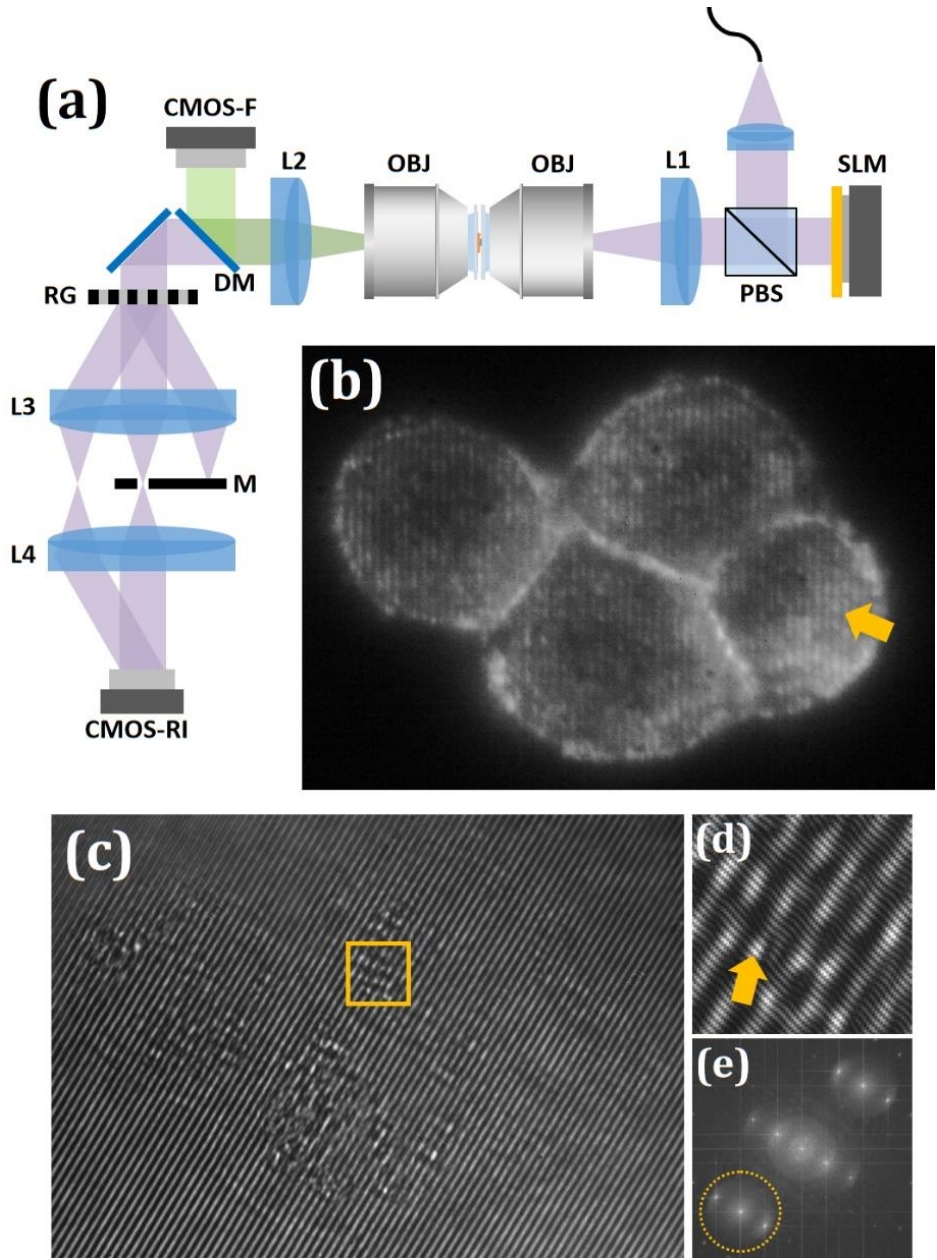

**Fig. 2. (a)** Experimental system enabling 3D fluorescence and refractive-index imaging. Examples of raw **(b)** fluorescent and **(c)** coherent-interferometric acquisitions required for SI reconstructions. **(d)** Zoom of interferogram shows both the SI and off-axis interference patterns. **(e)** Fourier-transform of interferogram shows sample's complex electric-field distribution.

## B. Raw image acquisition procedure

Though the presented optical system above contains detection arms for both coherent and fluorescent imaging, the SI patterns required to allow RI and super-resolution fluorescent reconstructions required different illumination procedures – thus coherent and fluorescent acquisitions were not simultaneous.

As in conventional DT, whole RI volumes were simply reconstructed with a dense angular sampling of the object's scattering potential, and required no axial scanning of the object. 32 increments of sinusoidal illumination spatial-frequency magnitude, within the range [0, NA/λ], were used per illumination rotation, with six total rotations spaced π/3 radians apart. For each spatial frequency, 5 translations of the pattern, at 2π/5 radian increments, were acquired. This resulted in 960 raw acquisitions to reconstruct a single RI volume. The camera integration time was set to 15 ms, thus the total acquisition time for RI imaging was under 15 seconds.

As explained in previous sections, fluorescent super-resolution via SI does not require incrementing the spatial-frequency magnitude of the sinusoidal illumination pattern – however, the object does need to be axially scanned. In this work, the object was axially scanned through 12.8 um at increments of 200 nm, satisfying the Nyquist requirement set by our fluorescent axial diffraction limit. For each axial position, the object was illuminated with the maximum spatial-frequency supported by the illumination objective's NA, with 5 pattern translations per spatial-frequency, spaced 2π/5 radian increments. Each pattern underwent 2 rotations, spaced π/2 radians apart. Camera integration time was set to 150 ms per acquisition, thus the total acquisition time for 3D super-resolved SI fluorescent imaging was ~96 seconds.

## C. SI visualization of polystyrene microspheres
### *1) Sub-diffraction resolution in SI-enabled RI visualization*
Using the principle of SA, coherent microscopy achieved by angularly illuminating the object has been demonstrated to achieve resolutions beyond the $d_{c,\parallel} = \lambda_c/\text{NA}$ lateral coherent diffraction-limit [4-9]. As noted above, SI imaging theoretically equates to a multiplexed form of SA when operated in the coherent domain, and previous works have demonstrated that SI-enhanced visualizations of coherent scatter and quantitative-phase (QP) do indeed contain sub-diffraction resolution content [35-43]. Here, we extend these results to rigorously demonstrate that SI-enabled RI reconstructions also visualize the sample at sub-diffraction resolutions. We do this by comparing RI visualizations with those of widefield QP, which we treat as our comparison standard for diffraction-limited coherent imaging. We refer the reader to other works for rigorous demonstrations of SI's capabilities for sub-diffraction resolution fluorescent imaging [29-34].

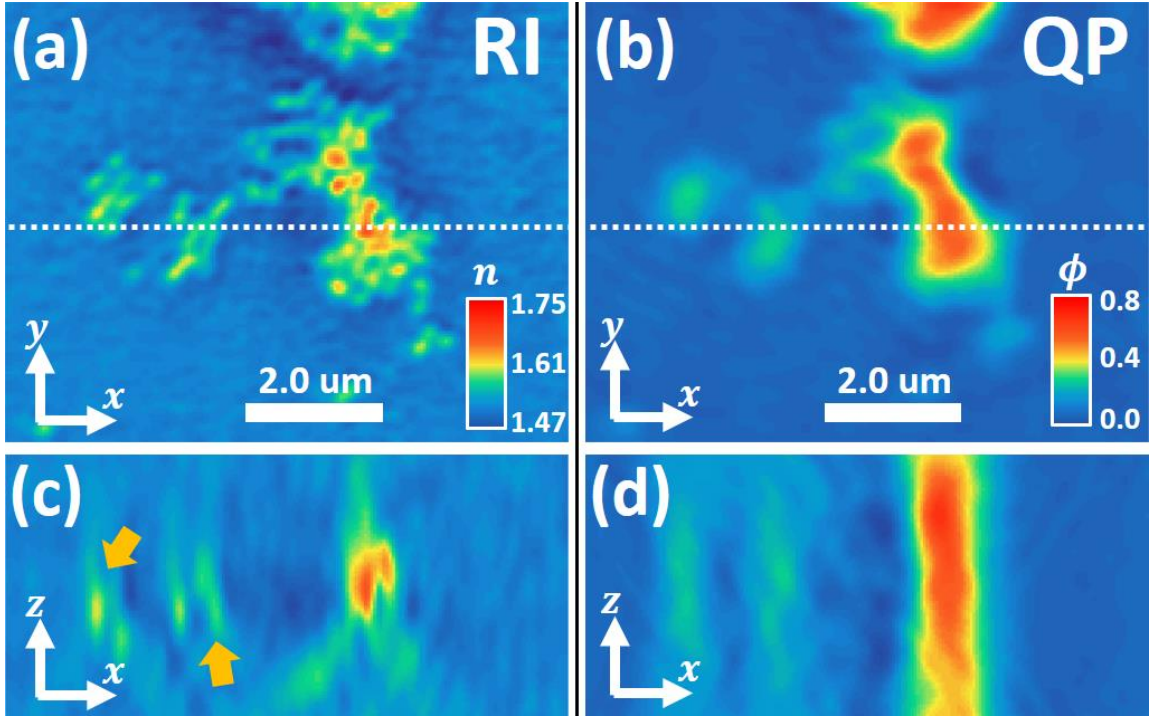

**Fig. 3. (a)** SI-enabled RI and **(b)** widefield diffraction-limited QP lateral visualizations of a cluster of 300 nm polystyrene microspheres. **(c)** Associated axial RI and **(d)** QP cross-sections are also shown. Lateral and axial visualizations are shown at identical size scales.

In Figs. 3(a,b), we show lateral visualizations of a 300 nm diameter polystyrene ($n(\lambda)$ = 1.605 at $\lambda$ = 488 nm) microsphere cluster immersed in glycerol ($nm(\lambda)$ = 1.47 at $\lambda$ = 488 nm) with SI-enabled RI and diffraction-limited QP contrast, respectively. As expected, given the system's coherent diffraction limit $d_{c,\parallel} = \lambda/\text{NA} \approx 350$ nm, QP imaging (Fig. 3(b)) shows no resolution of individual microspheres. In contrast, the RI reconstruction (Fig. 3(a)) is able to visualize individual microspheres. Cross-sectional axial cuts of the RI and QP volumes, through the region identified by the dashed white lines in Figs. 3(a,b), are also shown in Figs. 3(c,d), respectively. As mentioned above, diffraction-limited coherent imaging has no axial resolution [48, 49], and Fig. 3(d) demonstrates that QP imaging is unable to localize the sample in depth. Fig. 3(c), however, demonstrates that the RI reconstruction clearly visualizes the axial positions of individual microspheres, as indicated by yellow arrows.

## *2) SI-enabled multimodal 3D RI/fluorescent visualization*

We next demonstrate SI-enabled 3D dual-modal RI/fluorescent imaging on a sample where the fluorescent and RI signal outline identical physical structures. Though introduced in this work as simply proof-of-concept, this demonstration could serve as an important one-time calibration step for inter-modal registration in situations where the RI and fluorescent signals visualize independent components of a biological sample.

In Fig. 4, we demonstrate 3D RI/fluorescent visualizations of a monolayer of 2.0 um fluorescent polystyrene ($n(\lambda)$ = 1.605 at $\lambda$ = 488 nm) microspheres (BangLaboratories) immersed in index-matching oil (with calibrated $n_m(\lambda) = 1.545$ at $\lambda = 488$ nm). As expected, lateral and axial cross-sectional visualizations through the imaging volume (Figs. 4(a,b) and Figs. 4(c,d), respectively) demonstrate excellent RI/fluorescent agreement. We note that the RI axial cross-section contains RI signal (indicated by arrows in Fig. 4(c)) positioned above and below the axial plane of the microsphere monolayer – this artifact is a manifestation of the missing-cone problem prevalent in DT [52]. Though a positivity constraint was used in the RI reconstruction to computationally fill in the missing-cone region, such artifacts were not observed to fully disappear. Because the missing cone-problem is resolved in SI fluorescent super-resolution imaging [31], the SI fluorescent axial cross-section (Fig. 4(d)) shows the fluorescent signal to be well localized only to the plane of the microspheres. Figs. 4(e,f) show 3D visualizations of the SI-enabled RI and fluorescent volume reconstructions. Throughout this work, 3D renderings of volume reconstructions were done using Icy, an open-source biological image analysis platform.

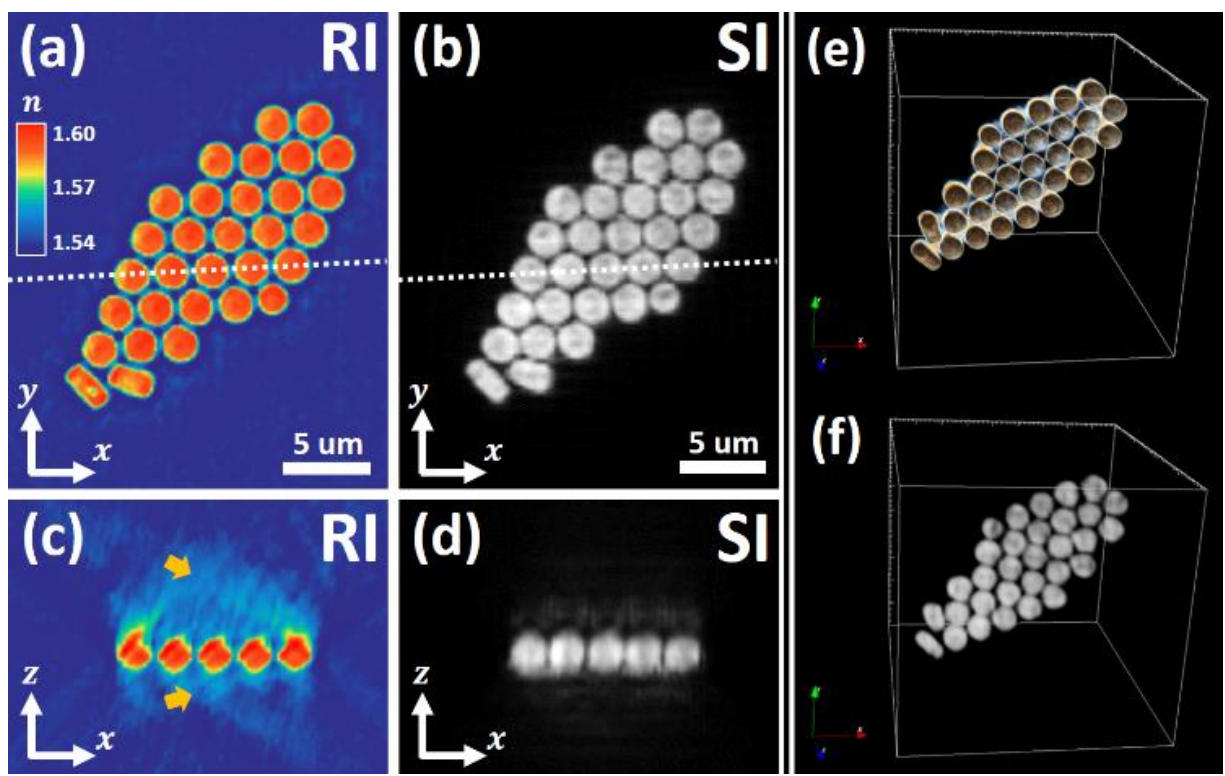

Fig. 4. SI-enabled (a) RI and (b) fluorescent lateral visualizations of a monolayer sample of 2.0 um fluorescently tagged polystyrene microspheres. Associated axial (c) RI and (d) fluorescent cross-sections are also shown. (e) RI and (f) fluorescent volume-renderings are also shown

## D. SI visualization in biological imaging
### *1) Dual-modality RI/fluorescent visualization of human alveolar basal epithelial adenocarcinoma*

We first demonstrate 3D RI/fluorescence sub-diffraction resolution biological imaging of adenocarcinomic human alveolar basal epithelial (A549) cells, fluorescently tagged for F-actin visualization (details on our experimental preparation procedures are presented in Supplement 1).

These cells are widely utilized as in-vitro models for Type II alveolar epithelial cells, and are frequently used in respiratory studies involving drug delivery and metabolism via the pulmonary route [53]. In Fig. 5, SI-enabled visualization capabilities for 3D RI and fluorescent volumetric imaging are illustrated for a single A549 cell. We start by illustrating a single plane from the reconstructed RI volume (Fig. 5(a)), a maximum-intensity axial projection through the reconstructed fluorescent super-resolution volume (Fig. 5(b)), and associated RI/fluorescent axial cross-sections from locations indicated by dashed white line in the lateral views, respectively. Fig. 5(a) demonstrates the A549 cell cytoplasm to have very similar RI to the surrounding media (n ~1.332), but to contain a large nucleus of higher RI (n ~1.35-1.36). Interestingly, these measured nuclear RI values are observed to match the values reported in a recent study measuring nuclear/cell-body RI across multiple cell-lines (though we urge readers to not generalize measurements from a single cell sample to whole cell populations, as justified below) [54]. The cell body is also shown to be scattered throughout with several high-density localizations with high RI (n > 1.39). We hypothesize these localizations to be either lipid-based vesicles [55], which are known to have high RI lipid bilayer, or lamellar bodies, which are known to be lipid storage and secretory units prevalent in Type II alveolar epithelial cells and responsible for the pulmonary surfactant system [56, 57]. Looking at the RI axial cross-section, these high-density localizations are relatively co-planar and positioned below the nucleus. Fig. 5(b) shows the F-actin cytoskeleton in the cell, and interestingly demonstrates higher F-actin density at the periphery of the region occupied by the cell's nucleus (identified from Fig. 5(a)), seemingly acting as a nuclear "pocket" of F-actin. This phenomenon was not generalized across other A549 cells from the same population (not shown here). Because the A549 cell cytoplasm demonstrates RI similar to that of the media, it is difficult to visualize the 3D shape of the cell body via RI contrast, as is evidenced by the axial RI cross-section in Fig. 5(a), which only visualizes the nucleus and high RI localizations and not the cell body's boundary. In contrast, F-actin cytoskeleton, known to encapsulate the general perimeter of the cell body, shows lateral and axial fluorescent views (Fig. 5(b)) that clearly illustrate the A549 cell body's generally flattened morphology. This observation matches the cell's general classification as a squamous epithelial cell [58]. Figs. 5(c,d) show 3D renderings of the reconstructed RI and fluorescence volumes, respectively. Figs. 5(e-g) and Figs. 5(h-j) show z-planes axially incremented through the RI and diffraction-limited QP volume reconstructions (QP volume was generated by digitally propagating a single planar QP electric-field). As observed in our previous works, QP shows little axial sectioning [42, 44].

Analogously, z-planes through the SI super-resolved (Figs. 5(k,m,o)) and diffraction-limited widefield (Figs. 5(l,n,p)) fluorescent volumes are also compared. As is evident, without SI super-resolution enhancement, the z-planes from the widefield fluorescent volume show copious amounts of defocused fluorescent signal, which obstructs in-focus visualization F-actin.

We emphasize that the RI values reported above were observed for a single A549 cell. Thus, these values should not be immediately generalized to A549 cell lines without consideration of multiple factors, such as temperature, plated-density, preparation procedures, stage of cycle during which the cell was imaged, and intrinsic natural variation of intracellular RI [8, 26, 59, 60]. We illustrate this issue in Supplement 1, where we show examples of cells imaged from the same cell population that have dramatically different RI distributions. It is conceivable that different research studies that analyze cell samples from the same cell line but under varying experimental conditions may report RI parameters/distributions with even more variation.

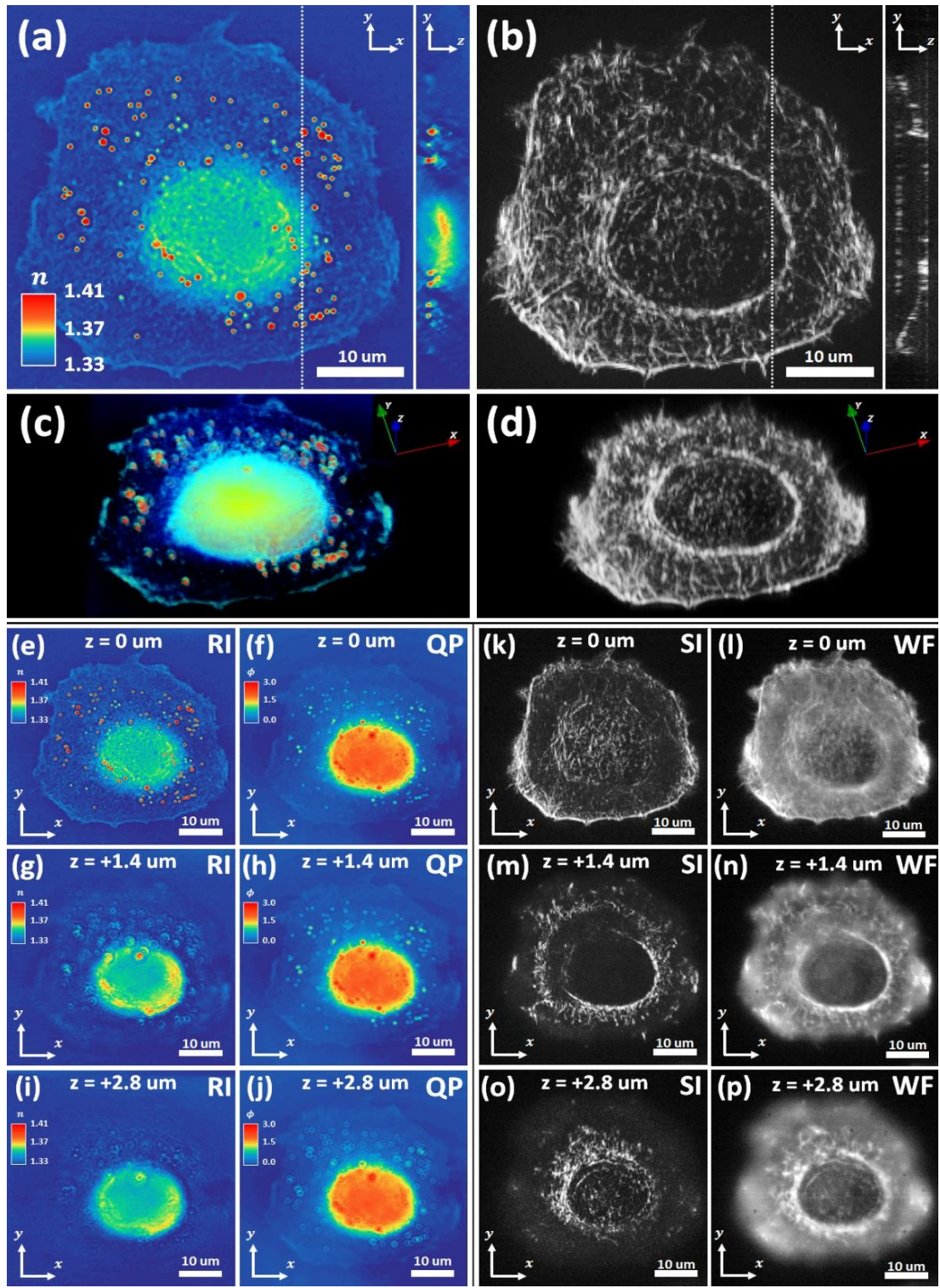

**Fig. 5.** Three-dimensional **(a)** RI and (b) fluorescence volumes are shown after SI reconstruction for a single A549 cell. (c,d) Corresponding 3D RI and fluorescent tomograms are shown [61]. Cross-sectional z-planes are compared through the (e,g,i) RI and (f,h,j) general QP volume reconstructions. Analogously, cross-sectional z-planes are also compared between the (k,m,o) SI-enabled super-resolution and (l,n,p) diffraction-limited widefield fluorescent volumes.

## 2) Dual-modality RI/fluorescent visualization of human colorectal adenocarcinoma

We next demonstrate biological RI/fluorescent 3D reconstruction of human colorectal adenocarcinoma (HT-29) cells, also fluorescently tagged for F-actin visualization (details on our experimental preparation procedures are presented in Supplement 1). Popularly used in studies exploring differentiation and function of intestinal cells, as well studies of tumor biology [62, 63], HT-29 cells are characteristically globular in shape and exhibit rounded cross-sectional profiles [8, 64]. In Fig. 6 below, we demonstrate 3D RI and fluorescent sub-diffraction reconstruction of three conjoined HT-29 cells. Immediately from Fig. 6(a), we see that the HT-29 cells compositionally contain greater RI ($n > 1.34$) than the surrounding media ($n \sim 1.33$), and are thus clearly outlined with RI visualization, unlike the A549 cell in Fig. 5. The HT-29 cell cytoplasm RI ($n \sim 1.34 – 1.355$) matches well with reports from previous work, notwithstanding minor differences due to dispersion [8, 59].

The HT-29 cell outlines are well matched by the F-actin visualization (Fig. 6(b)), as expected. Similar to the A549 cells, F-actin encapsulates the perimeter of the HT-29 cells and demonstrates a globular cross-section (Fig. 6(d)). Also similar to the A549 cells, the HT-29 cells demonstrate high-density RI localizations – however, these localizations are not co-planar (Fig. 3(c)) nor randomly scattered throughout the cell body, instead congregating towards the perimeter. Also in contrast to the A549 cells, there is no clear distinction in RI values separating the HT-29 nuclei from the cell bodies. Similar observations have been reported in previous studies of RI tomography [8, 44, 64]. More interestingly, however, is that that regions adjacent to the nuclear periphery (indicated by red arrows in Fig. 6(a) and identified by comparisons to past works) exhibit moderate RI ($n \sim 1.36$) and visually appear as indistinct haze. Furthermore, though nucleoli (indicated by yellow arrows in Fig. 6(a)) boundaries are visualized, the nucleoli interiors seem to match the cell body in RI. These observations of the nuclear periphery and cell nucleoli for this specific set of HT-29 cells are at odds with the observations of the cluster of HT-29 cells visualized in our previous study (Fig. 8 in [44]), where the nuclear periphery and cell nucleoli demonstrated high RI and distinct boundaries. We hypothesize that this observation indicates that the cluster of HT-29 presented in this study may have entered cell division and have neared the end of prophase, which is marked by a breakdown of the nuclear envelope and endoplasmic reticulum as well as the dissolution of the nucleolus.

Figs. 6(e-g) and Figs. 6(h-j) compare the SI-enabled RI visualizations to the coherent diffraction-limited QP visualizations for selected z- planes through the sample, respectively (for the region indicated by dashed yellow rectangle in Fig. 6(a)). As observed in our previous study [44], the high RI localizations show clear axial sectioning and variation in the RI imaging volume while defocus artifacts are abundant in the QP volume. Similarly in the fluorescent case, F-actin morphology shows clear development with respect to axial position when visualized after SI fluorescent super-resolution (Figs. 6(k-m)). In contrast, diffraction-limited fluorescent images show abundant out-of-focus signal in all axial slices (Figs. 6(n-p)). 3D tomograms are rendered to more intuitively visualize the HT-29 cluster volume using RI contrast (Figs. 6(q-s)) and super-resolution fluorescence (Fig. 6(t)). Figs. 6(q,r,s) show the RI volumetric visualizations thresholded at low ($n > 1.332$), medium ($n > 1.35$), and high ($n > 1.375$) RI to emphasize the cell's main body, other intracellular components, and high RI components, respectively.

As noted in the preceding section, these reported RI values for the HT-29 cell cluster should not be generalized to the whole HT-29 cell line without consideration of experimental conditions and internal intra-cell dynamics.

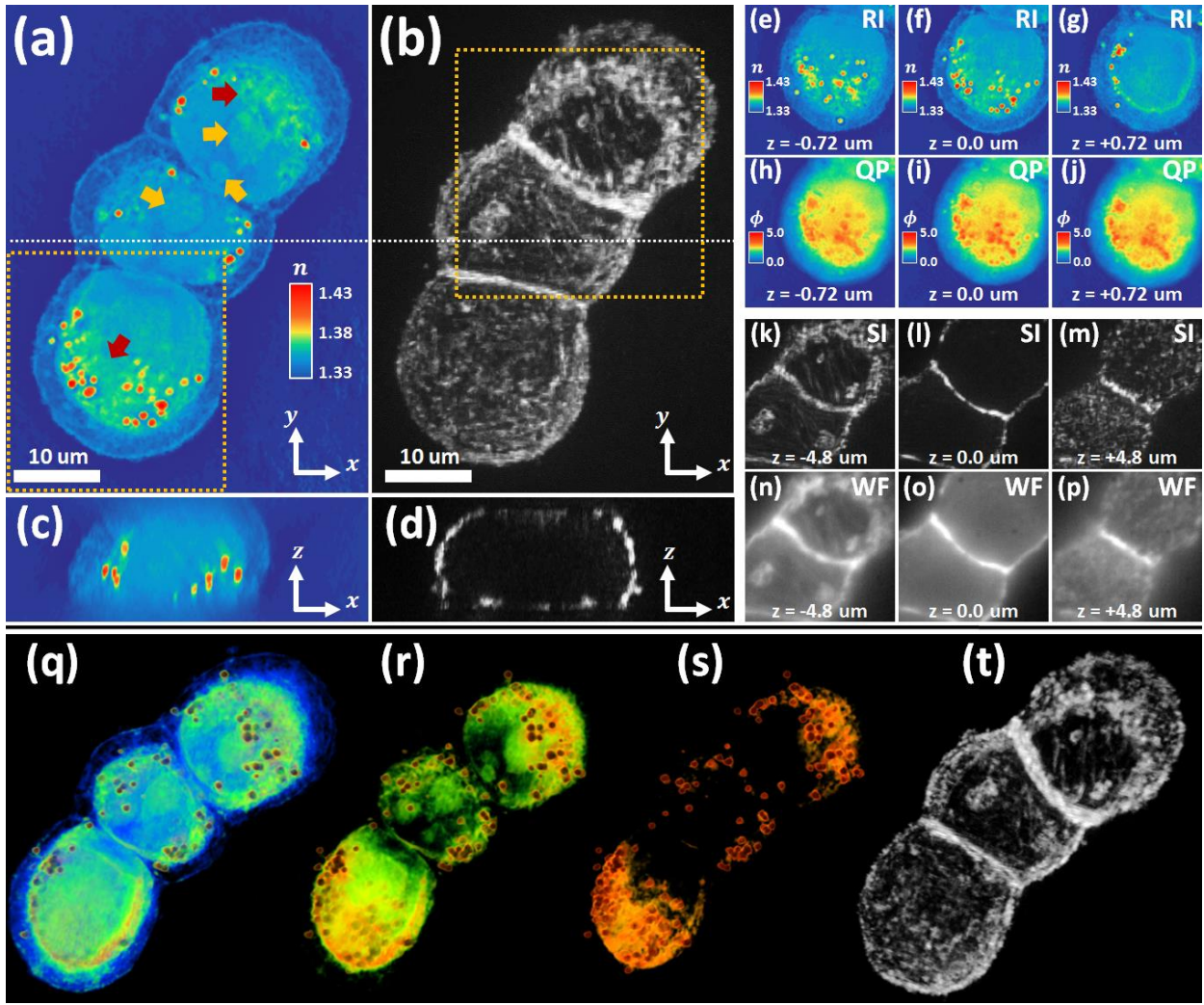

**Fig. 6. (a)** RI slice and **(b)** fluorescence maximum intensity projection shown for a HT-29 cell cluster, along with **(c,d)** associated axial cross-sections, respectively. Red and yellow arrows in (a) indicate nuclear periphery and nucleoli boundaries, respectively. From the coherent arm of the optical system, **(e-g)** RI and **(h-j)** standard QP visualization is compared for the cell cluster at select axial positions. Similarly, for the fluorescent arm of the system, fluorescent **(k-m)** super-resolution and **(n-p)** diffraction-limited widefield visualization of the cell cluster is also compared. 3D Tomograms are visualized with **(q-s)** different RI-thresholded opacity constraints (to visualize different intracellular components of varying RI) as well as with **(t)** general super-resolved fluorescent contrast

# IV. Discussion and conclusion

Though typically associated with fluorescent super-resolution, SI microscopy has been shown by recent works to also have applications toward sub-diffraction resolution coherent imaging. Because coherent SI essentially equates to a multiplexed form of SA, these sub-diffraction coherent capabilities derive from the same optical mechanisms that are central to angle-scanning DT. Thus, compared to typical fluorescent super-resolution or coherent SA techniques, SI is uniquely suited to enable both coherent and fluorescent visualizations at sub-diffraction resolutions. Following this motivation, we have experimentally demonstrated here, for the first time to our knowledge, the SI microscopy framework adapted into a multimodal RI/fluorescent 3D sub-diffraction resolution imaging solution.

We believe this work demonstrates that SI may have important implications in a host of studies that analyze separate, but synergistically operating, biological interactions. Specifically positioned to benefit from the work presented here are the biological studies requiring significant biochemical/biophysical and molecular analysis. Coherent imaging is a popular imaging choice when seeking to optically probe the quantitative biochemical and biophysical parameters of biological samples, and has been used to study cell mass, mechanics, spectroscopy, shear stress, etc [8, 18-21]. Furthermore, such studies do not require cells to undergo major preparation steps, and are targeted towards endogenous contrast. Important biological applications of RI imaging specifically have included quantitative extraction of cellular hemoglobin concentrations, volume/mass parameters, and deformability properties as well as endogenous visualization of cell dynamics [26-28]. However, such visualizations and parameter extractions have no molecular specificity, and thus are typically limited to studies of whole-cell analysis. In contrast, fluorescence imaging is the standard for molecular-specific visualization, and has played an important role in studying organelle dynamics, diffusion kinetics, intracellular transport, gene expression, etc [13-17]. In both cases of coherent and fluorescent imaging, the improved visualization from addition of sub-diffraction resolution has shown promise in enhancing biological insight. Unfortunately, due to the divide between coherent and fluorescent sub-diffraction resolution techniques, such insights are typically limited to one modality or the other. In regards to this issue, we believe that SI is uniquely positioned as an efficient imaging solution to enable comprehensive and synergistic analysis of biological components/interactions that could previously only be studied separately. Specifically, we believe that important applications of SI include studies into molecular-specific RI at sub-diffraction resolutions. Such studies could efficiently enable exploration into the structural, mechanical, and biochemical properties of individual molecularly-specific biological components and their relationships in molecular processes, dynamics, and functions.

**Funding.** National Science Foundation (NSF) (1403905);

**Acknowledgment.** We thank Brianna Loomis for help in preparation of biological samples and other members of the Izatt lab for useful discussions.

# Supplementary Information

In this section, we outline the procedures undertaken to prepare the calibration/biological samples used in our demonstration for SI-enabled refractive-index (RI)/fluorescent sub-diffraction resolution reconstructions. We show biological examples to emphasize that optical-path-length (OPL) and RI are coupled into quantitative-phase (QP) images, and that using only QP to infer biological RI may lead to erroneous conclusions. Furthermore, we also demonstrate that intracellular RI, apart from having a natural distribution across cells (even cells from the same cell-line), may change drastically throughout the cell cycle.

## 1. Sample preparation for calibration and biological imaging

*a) Coherent propagation of spatial frequencies through system aperture*

Calibration samples of 300 nm and 2.0 um diameter fluorescent polystyrene microspheres ($n(\lambda)$ = 1.605 at $\lambda$ = 488 nm, BangLaboratories) were prepared. 20 uL of microsphere dilutions (10 uL stock-solution/300 uL isopropyl alcohol) were placed onto #1.5 coverslips and allowed to dry, which facilitated the formation of microsphere monolayers. Glycerol (nm($\lambda$) = 1.47 at $\lambda$ = 488 nm) and high-index oil ($n$m($\lambda$) = 1.545 at $\lambda$ = 488 nm) were used to index-match the 300 nm and 2.0 um microspheres, respectively. An adhesive spacer and a second #1.5 coverslip were placed on top of the sample to assure a uniform sample layer for optimal high-NA imaging. Immersion oil (n($\lambda$) = 1.51 at $\lambda$ = 488 nm) was placed below and above the first and second #1.5 coverslips as the appropriate immersion media for the high NA illumination and detection microscope objectives (63X, 1.4 NA, Zeiss), respectively.

*B. Sample preparation for A549 and HT-29 biological cells*

A549 lung cancer cells and HT-29 colon cancer cells were imaged to demonstrate the multi-modal RI and fluorescence capabilities. HT-29 cells were grown in McCoy's 5A Medium supplemented with 10% fetal bovine serum (FBS). A549 cells were grown in high glucose Dulbecco's Modified Eagle Medium (DMEM) supplemented with 10% FBS. Both cell types were plated onto #1.5 coverslips at low-density to image isolated cells. After attaching overnight, a 4% paraformaldehyde in phosphate buffer solution (PBS) was used to fix the cells. Next, cells were permeabilized with a 0.1% Triton X-100 solution and later stained with Alexa Fluor 488 phalloidin (Life Technologies) for F-actin visualization. Following staining, the cells were washed with PBS. An adhesive spacer was placed around the cells in order to encapsulate the imaging volume between two #1.5 coverslips. Immersion oil (n($\lambda$) = 1.51 at $\lambda$ = 488 nm) was placed below and above the first and second #1.5 coverslips as the appropriate immersion media for the high NA illumination and detection microscope objectives (63X, 1.4 NA, Zeiss), respectively.

## 2. RI reconstruction separates cellular thickness from optical density

It is tempting to form conclusions about a sample's physical thickness or optical density (measured with RI) by simply considering the sample's QP distribution, which is often visualized topographically. In such situations, it is important to recall that QP is a measurement of optical-path-length (OPL), which is mathematically an accumulation of sample RI across the sample's physical thickness. Thus, sample RI and physical thickness cannot be individually extracted from a QP map without prior assumptions of the physical thickness or RI, respectively. Such assumptions have in fact been quite popular in a host of quantitative imaging studies, and applications range from using QP to extract biological RI under specific assumptions of sample morphology, to using QP for visualizations/measurements of cellular thickness during cellular growth and dynamics with prior knowledge of RI [1-4]. In cases where such assumptions are well substantiated in biology, QP may indeed be an effective solution for quantitative biological analysis. However, in cases where such assumptions are not well met, biological image analysis of QP may yield erroneous conclusions about sample RI or thickness.

To illustrate this point, we show in Fig. S1 a comparison example between RI and QP 3D visualization of a set of two conjoined A549 cells. RI and QP lateral visualizations of the cells, as shown in Figs. S1(a,b) respectively, both show the general outline of the cell cluster, as well as the locations of the high-RI localizations and cell nuclei/nucleoli. However, it is important to note that Fig. S1(b) demonstrates that the cell nuclei and nucleoli, with QP signals of ~1.6 rad and ~2.1 rad respectively, have similar OPL to the high-RI localizations scattered throughout the cell bodies. In contrast, Fig. S1(a) demonstrates that these localizations have significantly higher RI ($n > 1.40$) than the cell nuclei ($n > 1.35$) or nucleolus ($n > 1.36$). This implies that the high-RI localizations must have significantly smaller physical thicknesses than the cell nuclei/nucleolus. Fig. S1(c) confirms this by showing a RI axial cross-section across the dashed white line indicated in Fig. S1(a). As is evident, the physical thicknesses of the high-RI localizations are significantly smaller than the thicknesses of the nuclei (indicated with yellow arrow in Fig. S1(c)) and nucleoli (indicated with red arrows in Fig. S1(c)). Thus, though QP imaging may allow one to observe that cell nuclei/nucleoli and high-RI localizations have similar OPL, extrapolating this observation to conclude that such components have similar optical densities or physical thicknesses would be erroneous. Hence, RI reconstruction allows decoupling of physical thickness and optical density from OPL, and in this example, yields biologically relevant information about the sample that cannot be obtained from QP imaging.

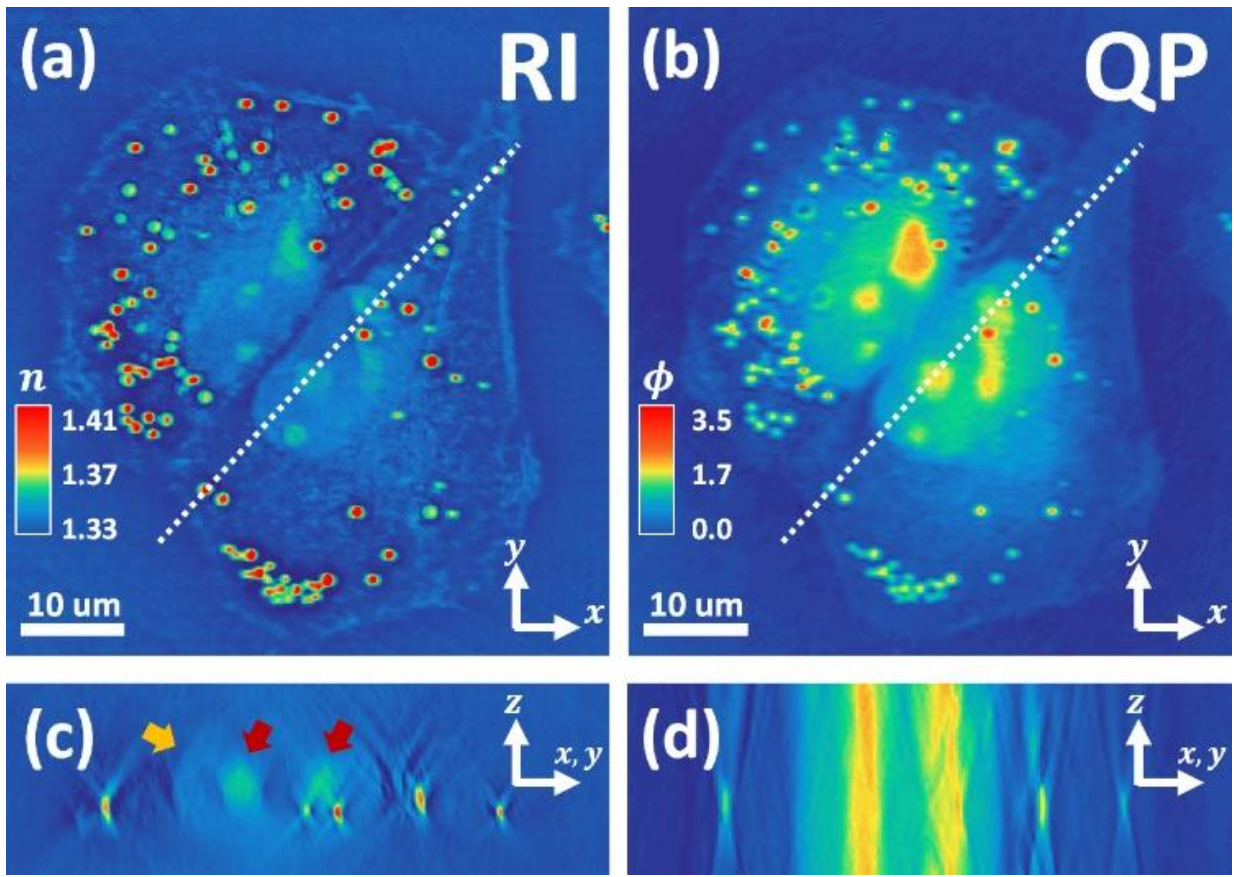

**Fig. S1. (a)** SI-enabled RI and **(b)** widefield diffraction-limited QP lateral visualizations of a conjoined pair of A549 cells. **(c)** Associated axial RI and **(d)** QP cross-sections are also shown. 3D QP signal was reconstructed by using Fresnel kernels to digitally propagate the in-focus QP image in (b).

## 3. Intra-cellular RI distribution naturally varies within a cell population

Though we have demonstrated here that structured illumination can quantitatively reconstruct RI alongside super-resolution fluorescence, we emphasize that biological conclusions based on quantitative measurements must be distinguished between the singe cell and cell population levels. In the main text of this work, we described the RI distributions of selected A549 and HT-29 cells (Figs. 5 and 6 in the main text) – however, we emphasize that such descriptions, which were based on individual cells, should not be rigorously generalized to whole cell populations. Due to RI distributions naturally varying across individual cells even from the same population, population level conclusions are best made with statistical significance, which requires observation of several cells. Furthermore, intracellular RI has been demonstrated to be dependent on cell stage, temperature, etc – hence, care should be taken in ensuring that cells are from matching conditions before being included in a composite RI tabulated dataset [2, 5-7].

To illustrate how drastically RI distributions may vary between cells from even the same population, we show in Fig. S1 comparisons between two cells from A549 and HT-29 cell populations, in Figs. S2(a,b) and Figs. S2(c,d), respectively (Figs. S2(a,c) illustrate the A549 and HT-29 cells introduced in Figs. 5 and 6 in the main text, respectively). Comparing within the A549 cell line, we see that Sample #1 (Fig. S2(a)) and Sample #2 (Fig. S2(b)) visualize cells with nuclei that are dramatically different in size and optical density. Sample #2 visualizes a conjoined pair of A549 cells (same sample shown in Fig. S1) that demonstrate cell nuclei (indicated with yellow arrows in Fig. S1(b)) with lower RI than in Sample #1. Furthermore, nucleoli appear fully formed (indicated with red arrows in Fig. S1(b)) and demonstrate moderately higher RI compared to the nuclei. In contrast, Sample #1 does not demonstrate distinct nucleoli boundaries within its nucleus, which is considerably enlarged and heterogeneously distributed with RI higher than in the nuclei of Sample #2. We hypothesize that this may be due to nucleoli breakdown, which resulted in the release of high-RI content into the nucleus. Combined with the observation of large nuclei size, we hypothesize that Sample #1 may be visualizing a cell undergoing preparation to enter cell division [8].

When comparing within the HT-29 cell line, we see that both Sample #1 (Fig. S2(c)) and Sample #2 (Fig. S2(d)) visualize a 3-cell cluster. However, the cells in Sample #2 show clear protrusions of the cell membrane (indicated by yellow arrows in Fig. S2(d)). These "blebs" are associated with expansion and eventual rupture of the cell membrane, and are common features of cell apoptosis. Furthermore, Sample #2 also shows a clear increase in number and size of high-RI localizations, compared to Sample #1, which would have the effect of raising the mean RI of the cells. Though identity of these localizations without direct fluorescent staining is difficult to ascertain, we hypothesize these localizations to be mainly apoptotic bodies, which are associated with apoptosis and blebbing [9].

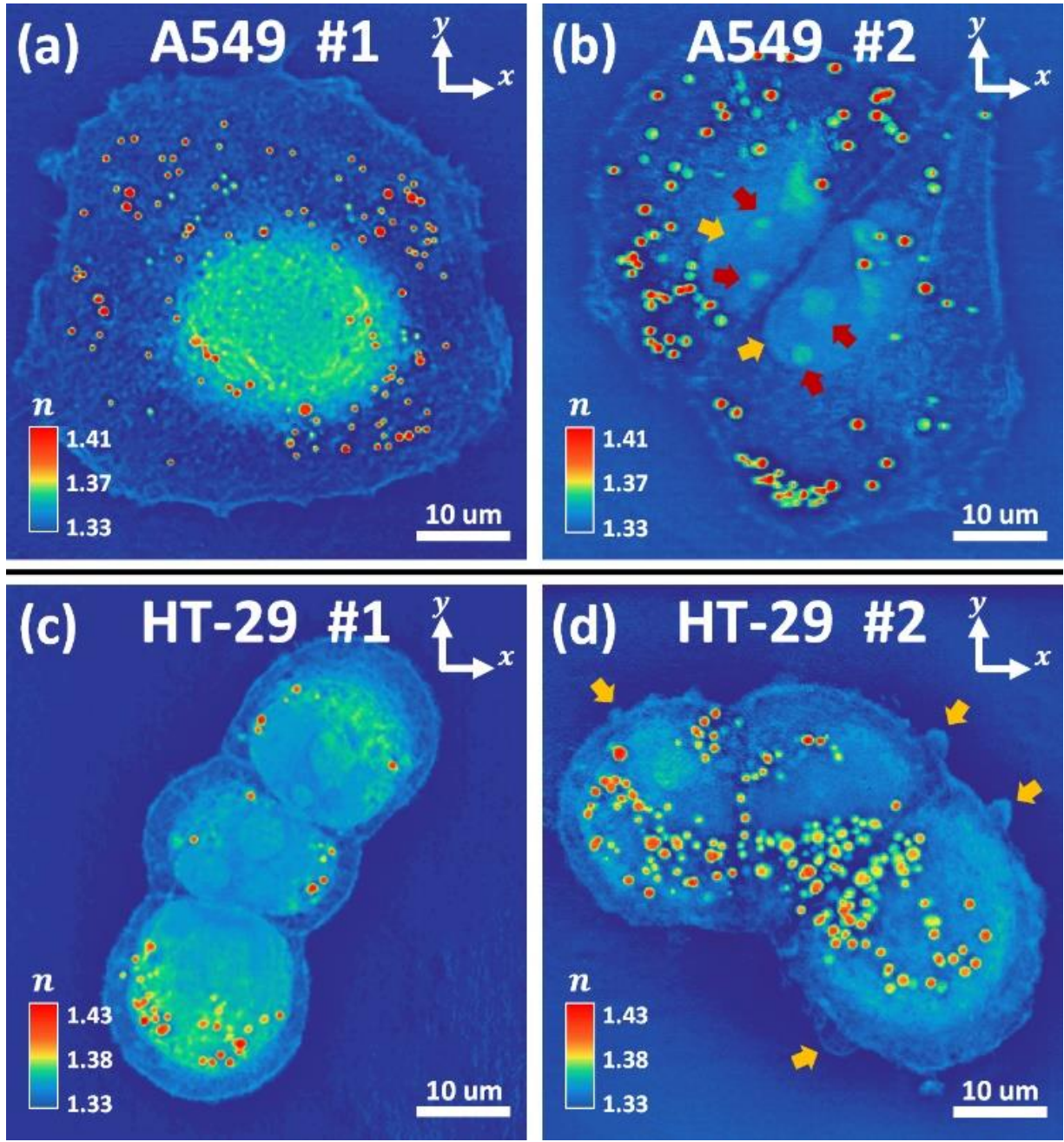

**Fig. S2.** Lateral visualizations of two samples of **(a,b)** A549 and **(c,d)** HT-29 cells. Yellow and red arrows in (b) indicate fully formed cell nuclei and nucleoli, respectively. Yellow arrows in (d) indicate cell blebs, an indication of apoptosis.

## 4. SI for RI/fluorescence dual-modality visualization of HT-29 blebbing

In Fig. S3(a,b) below, we demonstrate sub-diffraction resolution RI and fluorescence imaging, respectively, of a cluster of HT-29 cells (same sample shown in Fig. S2(d) ) that we hypothesize is undergoing apoptosis. As mentioned above, RI visualization of the cells demonstrates a large number of high-RI localizations as well as membrane blebbing. Fluorescent visualization also clearly reveals cellular blebbing, but also demonstrates increased disorder and sparse density of the F-actin cytoskeleton at locations near the blebs. As is evident, RI and fluorescence imaging visualize different structures that cannot be visualized with one modality alone. Hence, SI enables complementary multi-modal sub-diffraction resolution imaging, which may be important to cohesively study biological processes with important molecular and biophysical/biochemical components.

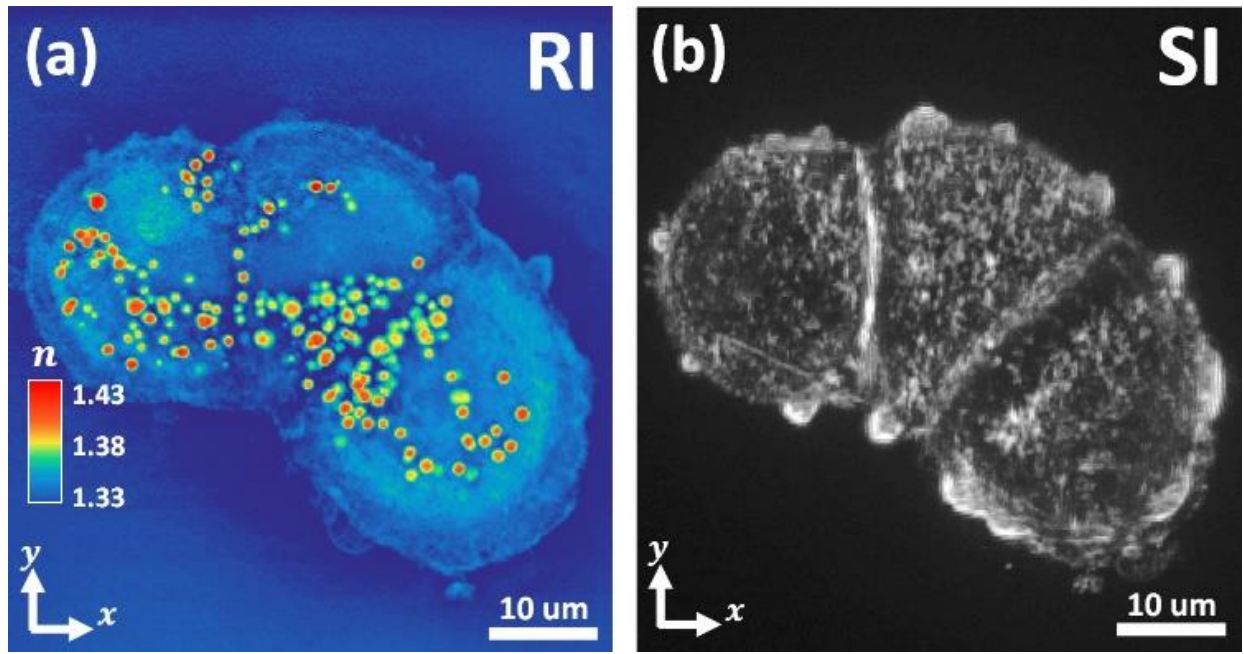

**Fig. S3.** Lateral visualizations of a 3-cell cluster of HT-29 cells undergoing apoptosis. Lateral **(a)** RI and **(b)** fluorescence visualizations are shown.